\documentclass[twocolumn,showpacs,preprintnumbers,amsmath,amssymb]{revtex4}

\usepackage{graphicx}
\usepackage{dcolumn}
\usepackage{bm}

\newcommand{\PP}{{\mathbb P}}
\newcommand{\TT}{{\mathbb T}}
\newcommand{\EE}{{\mathbb E}}

\newcommand{\cA}{\ensuremath{\mathcal A}}

\newcommand{\cE}{\ensuremath{\mathcal E}}
\newcommand{\cF}{\ensuremath{\mathcal F}}

\newcommand{\cH}{\ensuremath{\mathcal H}}

\newcommand{\cL}{\ensuremath{\mathcal L}}
\newcommand{\cM}{\ensuremath{\mathcal M}}

\newcommand{\cT}{\ensuremath{\mathcal T}}

\begin{document}


\title{Stationary non-equilibrium properties for a heat conduction model}

\author{C\'edric Bernardin}
\affiliation{Universit\'e de Lyon, CNRS (UMPA)\\
Ecole Normale Sup\'erieure de Lyon,\\
46, all\'ee d'Italie,\\
 69364 Lyon Cedex 07 - France.}
 \email{cbernard@umpa.ens-lyon.fr}
\thanks{The author is very grateful to L. Bertini to explain him MFT theory and for his help in the computations and to S. Olla for very valuable discussions. He acknowledges the support of the French Ministry of Education through the ANR BLAN07-2184264 grant.}

\date{\today}

\begin{abstract}
We consider a stochastic heat conduction model for solids composed by N
interacting atoms. The system is in contact with two heat baths at different temperature
$T_\ell$ and $T_r$. The bulk dynamics conserve two quantities: the energy and the deformation
between atoms. If $T_\ell \neq T_r$, a heat flux takes place in the system. For large $N$, the system adopts a linear temperature profile between $T_\ell$ and $T_r$. We establish the hydrodynamic limit for the two conserved quantities. We introduce the fluctuations field of the energy and of the deformation in the non-equilibrium steady state. As $N$ goes to infinity, we show that this field converges to a Gaussian field and we compute the limiting covariance matrix. The main contribution of the paper is the study of large deviations for the temperature profile in the non-equilibrium stationary state. A variational formula for the rate function is derived following the recent macroscopic fluctuation theory of Bertini et al.
\end{abstract}

\maketitle

\section{{\label{sec:intro}} Introduction}

Understanding of the steady state of non-equilibrium  systems is the subject of intense research. The typical situation is a solid in contact with two heat baths at different temperature. At the difference of equilibrium systems where the Boltzmann-Gibbs formalism provides an explicit description of the steady state, no equivalent theory is available for non-equilibrium stationary state (NESS). 

In the last few years, efforts have been concentrated on stochastic lattice gases (\cite{S1}). For these latter precious informations on the steady state like the typical macroscopic profile of conserved quantities and the form of the Gaussian fluctuations around this profile have been obtained (\cite{S1}). Recently, Bertini, De Sole, Gabrielli, Jona-Lasinio and Landim proposed a definition of non-equilibrium thermodynamic functionals via a macroscopic fluctuation theory (MFT) which gives for large diffusive systems the probability of atypical profiles (\cite{BL1},\cite{BL2}) in NESS. The method relies on the theory of hydrodynamic limits and can be seen as an infinite-dimensional generalization of the Freidlin-Wentzel theory. The approach of Bertini et al. provides a variational principle from which one can write the equation of the time evolution of the typical profile responsible of a given fluctuation. The resolution of this variational problem is, however, in general very difficult and it has only been carried for two models : the Symmetric Simple Exclusion Process (SSEP) (\cite{BL2}) and the Kipnis Marchioro Presutti (KMP) model (\cite{BL3}). Hence, it is of extreme importance to identify simple models where one can test the validity of MFT.  

The most studied stochastic lattice gas is the Simple Exclusion Process. Particles perform random walks on a lattice but jumps to occupied sites are suppressed. Hence the only interaction is due to the exclusion condition. The only conserved quantity by the bulk dynamics is the number of particles. In this situation, the heat reservoirs are replaced by particles reservoirs which fix the density at the boundaries. The KMP process is a Markov process composed of particles on a lattice. Each particle has an energy and a stochastic mechanism exchange energy between nearest-neigbor particles (\cite{KMP}). 

The real motivation is to extend MFT for Hamiltonian systems (\cite{B-r}). Unfortunately, for these later, even the derivation of the typical profile of temperature adopted by the system in the steady state is out of range of the actual techniques (\cite{BLR}). The difficulty is to show that the systems behave ergodically, e.g. that the only time invariant measures locally absolutely continuous w.r.t. Lebesgue measure are, for infinitely extended spatial uniform systems, of the Gibbs type. For some stochastic lattice gases it can be proven but it remains a challenging problem for Hamiltonian dynamics.

We investigate here  the MFT for a system of harmonic oscillators perturbed by a conservative noise (\cite{Ber1}, \cite{BO},\cite{BBO}). These stochastic perturbations are here to reproduce (qualitatively) the effective (deterministic) randomness coming from the Hamiltonian dynamics (\cite{OVY}, \cite{LO}, \cite{FFL}). This hybrid system can be considered as a first modest step in the direction of purely Hamiltonian systems. 

From a more technical point of view, SSEP and KMP are gradient systems and have only one conserved quantity. For gradient systems the microscopic current is a gradient (\cite{KL}) so that the macroscopic diffusive character of the system is trivial. Dealing with non-gradient models, we have to show that microscopically, the current is a gradient up to a small fluctuating term. The decomposition of the current in these two terms is known in the hydrodynamic limit literature as a \textit{fluctuation-dissipation equation} (\cite{EMY}).  In general, it is extremely difficult to solve such an equation. 

Our model has two conserved quantities, energy and deformation, and is non-gradient. But fortunately, an exact fluctuation-dissipation equation can be established. In fact we are not able to apply MFT for the two conserved quantities but only for the temperature field which is a simple, but non-linear, functional of the energy and deformation fields.

The paper is organized as follows. In section \ref{SEC:2}, we define the model. In section \ref{SEC:3} we establish the fluctuation-dissipation equation and obtain hydrodynamic limits for the system in a diffusive scale. Section \ref{SEC:4} is devoted to a physical interpretation of the fluctuating term appearing in the fluctuation-dissipation equation. In section \ref{SEC:5} we compute the covariance of the fluctuation fields in the NESS by a dynamical approach and show the covariance for the energy presents a non-locality we retrieve in the large deviation functional (the quasi-potential).  The latter is studied in section \ref{SEC:6} for the temperature field.       

\section{{\label{sec:2}} The Model}

We consider the dynamics of the open system of length $N$.  Atoms are labeled by $x \in \{1,\dots, N\}$. Atom $1$ and $N$ are in contact with two heat reservoirs at different
temperatures $T_{\ell}$ and $T_r$. Momenta of atoms are denoted by $p_1, \dots, p_{N}$ and the distance between particles are denoted by $r_1,
\dots, r_{N-1}$. The Hamiltonian of the system is given by
\begin{eqnarray*}
  {\cH}^N = \sum_{x=1}^{N} {e}_x, \quad  {e}_x = \frac{ p_x^2 + r_x^2}2 \qquad x= 1,\dots, N-1\\
  {e}_{N} = \frac{p_{N}^2}2.
\end{eqnarray*}
We consider stochastic dynamics where the probability
density distribution on the phase space at time $t$, denoted
by $P(t, p, r)$, evolves following the Fokker-Planck equation
\begin{equation*}
\partial_t P = N^2 {\mathcal L}^* P
\end{equation*}
Here ${\mathcal L} = {\mathcal A} + \gamma {\mathcal S}+{\mathcal B}_{1,T_\ell} +{\mathcal B}_{N,T_r}$ is the generator of the process and ${\mathcal L}^*$ the
adjoint operator. The factor $N^2$ in front of ${\mathcal L}^*$ is here because we have speeded up the time by $N^2$, this corresponds to a diffusive scaling. 

${\mathcal A}$ is the usual Hamiltonian vector field
\begin{eqnarray*}
{\mathcal A} = \sum_{x=1}^{N-1} (p_{x+1} - p_x) \partial_{r_x}
     + \sum_{x=2}^{N-1} (r_x - r_{x-1}) \partial_{p_x}\\ 
  + (r_1-\ell) \partial_{p_1} - (r_{N-1}-\ell) \partial_{p_{N}}
\end{eqnarray*}
The constant $\ell$ fix the deformation at the boundaries.

$\mathcal S$ is the generator of the stochastic perturbation and $\gamma>0$ is a positive parameter that regulates its strength. The operator $S$ acts only on momenta $\{p_x\}$ and generates a diffusion on the surface of constant kinetic energy. This is defined as follows. For every nearest neigbor atoms $x$ and $x+1$, consider the following one dimensional surface of constant kinetic energy $e$ 
\begin{equation*}
{\mathbb S}_e^1 =\{ (p_x,p_{x+1}) \in {\mathbb R}^2; p_x^2 +p_{x+1}^2 =e\}
\end{equation*} 
The following vector field $X_{x,x+1}$ is tangent to ${\mathbb S}_{e}^1$
\begin{equation}
 \label{eq:4}
   X_{x, x +1} = p_{x+1} \partial_{p_x} -  p_x \partial_{p_{x+1}}
\end{equation}
so $X_{x,x+1}^2$ generates a diffusion on ${\mathbb S}_e^1$ (Brownian motion on the circle).  We define
\begin{equation*}
{\mathcal S}= \frac {1}{2} \sum_{x=1}^{N-1} X_{x, x +1}^2 
\end{equation*}

${\mathcal B}_{1,T_\ell}$ and ${\mathcal B}_{N,T_r}$ are two boundary generators of Langevin baths at temperature $T_\ell$ and $T_r$
\begin{equation*}
{\mathcal B}_{x,T}= \frac 12 \left(T \partial_{p_x}^2 - p_x \partial_{p_x} \right)
\end{equation*}

The bulk dynamics conserve two quantities: the total energy ${\mathcal H}^N=\sum_{x=1}^{N} {e}_x$ and the total deformation ${\mathcal R}^N=\sum_{x=1}^{N-1} r_x$. The energy conservation law can be read locally as (\cite{Ber1}, \cite{BO})
\begin{equation*}
e_x (t) - e_x (0) = J^e_{x} (t) -J_{x+1}^e (t)
\end{equation*} 
where $J^e_{x} (t)$ is the total energy current between $x-1$ and $x$ up to time $t$. This can be written as 
\begin{equation*}
J^e_{x} (t)=N^2 \int_0^t j^e_{x} (s)ds + M_{x} (t)
\end{equation*}
In the above, $M_{x} (t)$ is a martingale, i.e. a stochastic noise with mean $0$. The instantaneous energy current $j^{e}_{x}$ can be written as
\begin{equation*}
j_{x}^e= -r_{x-1}p_{x} -\cfrac{\gamma}{2}\nabla(p_{x}^2) 
\end{equation*}
The first term $-r_{x-1} p_x$ is the Hamiltonian contribution to the energy current while the noise contribution is given by the discrete gradient $-(\gamma/2) \nabla(p_{x}^2)=(\gamma/2)(p_{x}^2 -p_{x+1}^2)$.

Similarly, the deformation instantaneous current $j^{r}_{x}$ between $x-1$ and $x$ is given by
\begin{equation*}
j^{r}_{x}=-p_{x}
\end{equation*}

We denote by $\mu^{ss}=<\cdot>_{ss}$ the invariant probability measure for the process. In the case $T_\ell = T_r=T$, the system is in thermal equilibrium. There is no heat flux and the Gibbs invariant measure (or canonical measure) is a product Gaussian measure $\mu_{ss}=\mu^{T,\ell}$ depending on the temperature $T$ and the mean deformation $\ell$:
\begin{equation}
\label{eq:mu}
\mu^{T,\ell} = Z_{T}^{-1} \exp \left\{ -\cfrac{1}{2T}\sum_{x=1}^{N} p_x^2 -\cfrac{1}{2T}\sum_{x=1}^{N-1} (r_x-\ell)^2  \right\}
\end{equation}

\section{{\label{sec:3}}  Fluctuation-dissipation equation and  Hydrodynamic limit}
Diffusive interacting particle systems can be classified in two categories: gradient systems and non-gradient systems (\cite{KL}). For the first, we can write the curent of the conserved quantities as a spatial discrete gradient. For example SSEP and KMP process are gradient systems.  A powerful approach introduced by Varadhan (\cite{V}) to study non-gradient systems is to obtain a fluctuation-dissipation equation, meaning a decomposition of the current $j$ of conserved quantities as the sum of a microscopic gradient $\nabla h$ and of a fluctuating term of the form ${\mathcal L} u$:
\begin{equation}
\label{eq:fde0}
j=\nabla h + {\mathcal L}u
\end{equation}
where ${\mathcal L}$ is the generator of the interacting particle system. In fact, the equality (\ref{eq:fde0}) is only an approximation in a suitable Hilbert space (\cite{KL}).    

Fortunately, for our system, we can write an equality like (\ref{eq:fde0}) without approximations.
The fluctuation-dissipation equation for the deformation current $j^r$ and the energy current $j^e$ is given by (\cite{Ber1})
\begin{equation}
\label{eq:fde}
\begin{cases}
j^{r}_{x}= -\gamma^{-1}\nabla(r_x) + {\mathcal L} h_x\\
j^e_{x}=\nabla\left[\phi_x \right]+{\mathcal L} g_x
\end{cases}
\end{equation}
where
$$\phi_x = \cfrac{1}{2\gamma}r_x^2 + \cfrac{\gamma}{2} p_x^2 +\cfrac{1}{2\gamma} p_x p_{x+1} +\cfrac{\gamma}{4}\nabla(p_{x+1}^2)$$
and
$$h_x= \gamma^{-1} p_x, \quad g_x =\cfrac{p_x^2}{4}+\cfrac{p_x}{2\gamma} (r_x + r_{x-1})$$

Assume that initially the system starts from a local equilibrium $<\cdot>$ with macroscopic deformation profile $u_0 (q)$ and energy profile ${\varepsilon}_0 (q)$, $q \in [0,1]$. This means that  if the macroscopic point $q \in [0,1]$ is related to the microscopic point $x$ by $q=x/N$ then at time $t=0$
\begin{equation*}
<r_{[Nq]} (0)> \to u_0(q), \quad <e_{[Nq]} (0)> \to {\varepsilon}_0 (q) 
\end{equation*}
as $N$ goes to infinity. The currents are related to conserved quantities by the conservation law
\begin{eqnarray*}
\partial_t <r_{[Nq]}(t)> \approx -N \partial_q <j_{[Nq]}^r (t)>,\\
\partial_{t} <e_{[Nq]}(t)> \approx - N \partial_q <j_{[Nq]}^e (t)>.
\end{eqnarray*}
By (\ref{eq:fde}) and the fact that the terms  $N<{\mathcal L} h_x>$ and $N<{\mathcal L} g_x>$ are of order ${\mathcal O} (N^{-1})$ and do not contribute  to the limit (\cite{Ber1}) we get 
\begin{equation*}
\begin{cases}
\partial_t <r_{[Nq]} (t)> \approx \gamma^{-1} \Delta <r_{[Nq]}(t)>\\
\partial_t <e_{[Nq]} (t)> \approx \Delta <\phi_{[Nq]} (t)>\\
\end{cases}
\end{equation*}  
To close the hydrodynamic equations, one has to replace the term $<\phi_{[Nq]} (t)>$ by a function of the conserved quantities $<r_{[Nq]} (t)>$ and $<e_{[Nq]}(t)>$. 

The replacement is obtained through a "thermal local equilibrium" statement (see \cite{Ber1}, \cite{BO}, \cite{ELS1},\cite{ELS2}, \cite{KL}, \cite{S1}) for a rigorous justification in the context of conservative interacting particle systems). We repeat here the arguments of \cite{BL3} for convenience of the reader. Thermal local equilibrium assumption corresponds to assume that each given macroscopic region of the system is in equilibrium, but different regions may be in different equilibrium states, corresponding to different values of the parameters. Let us consider an atom with position $q=x/N$ which is far from the boundary and introduce a very large number $2L+1$ of atoms in microscopic units ($L\gg 1$), but still an infinitesimal number at the macroscopic level ($(2L+1)/N \ll 1$). We choose hence $L=\epsilon N$ where $\epsilon \ll 1$ in order to have these two conditions. We consider the system in the box $\Lambda_L (x)$ composed of the atoms labeled by $x-L, \ldots, x+L$. The time evolution of the $2L+1$ atoms is essentially given by the bulk dynamics; since the variations of deformation and energy in the volume containing the $2L+1$ atoms changes only via boundary effects and we are looking at what happened after $N^2$ microscopic time units, the system composed of the $L$ atoms has relaxed to the micro-canonical state $\lambda_{{\bar r}_q (t), {\bar e}_q (t)}$ corresponding to the local empirical deformation ${\bar r}_q (t)$ and the local empirical energy ${\bar e}_q (t)$ in the box $\Lambda_L (x)$. This means that we can divide the observables into two classes, according to their relaxation times: the fast observables, which relax to equilibrium values on a time scale much shorter than $t$ and will not have any effect on the hydrodynamical scales and the slow observables which are locally conserved by the dynamics and need much longer times to relax. We can then replace the term $<\phi_{[Nq]} (t)>$ by $\lambda_{{\bar r}_q (t), {\bar e}_q (t)} (\phi_0)$. By equivalence of ensembles, in the thermodynamic limit $N \to \infty$ and then $\varepsilon \to 0$, this last quantity is equivalent to
\begin{equation*}
\cfrac{\gamma+\gamma^{-1}}{2} <e_{[Nq]} (t)> + \cfrac{\gamma^{-1}-\gamma}{4} (<r_{[Nq]} (t)>)^2
\end{equation*}
We have obtained the time evolution of the deformation/energy profiles $u(t,q)=\lim <r_{[Nq]} (t)>$, $\varepsilon (t,q)= \lim <e_{[Nq]} (t)>$ in the bulk. At the boundaries, Langevin baths fix temperature at $T_\ell$ and $T_r$. Hence it is more natural to introduce the couple of deformation/temperature profiles rather than deformation/energy profiles. The temperature profile $T(t,q)$ is related to $u(t,q)$ and $\varepsilon (t,q)$ by $\varepsilon (t,q) = T(t,q) +u(t,q)^2/2$. Deformation and temperature profiles evolve according to the following equations
\begin{equation}
\label{eq:hl}
\begin{cases}
\partial_t T = \cfrac{1}{2}(\gamma+\gamma^{-1}) \Delta T + \gamma^{-1}(\nabla u)^2,\\
\partial_t u =\gamma^{-1} \Delta u,\\
T(t,0)=T_\ell,\;\; T(t,1)=T_r,\\
u(t,0)=u(t,1)=\ell,\\
T(0,q)=T_0 (q),\; u(0,q)=u_{0}(q).
\end{cases}
\end{equation}

As $t$ goes to infinity, the system reaches its steady state characterized in the thermodynamic limit by a linear temperature profile ${\bar T}(q)=T_\ell + (T_r -T_\ell) q$ and a constant deformation profile ${\bar r}(q)= \ell$. The system satisfies Fourier's law and the conductivity is given by $(\gamma+\gamma^{-1})/2$ (\cite{BO}).

\section{{\label{sec:4}} Interpretation of the fluctuation-dissipation equation}

We have seen that functions $h_x$ and $g_x$ had no influence on the form of the hydrodynamic equations. This is well understood by the fact that they are related to \textit{first order} corrections to local equilibrium as we explain below.

Assume $T_{\ell (r)}=T \pm \delta T /2$ with $\delta T$ small. For $\delta T = 0$, the stationary state $<\cdot>_{ss}$ equals the Gibbs measure $\mu_{T}^{\ell}$ (see \ref{eq:mu}). If $\delta T$ is small, it is suggestive to try an ansatz for $<\cdot>_{ss}$ in the form:
$$\tilde{\mu}  = Z^{-1} \,\prod_x dp_x dr_x \exp\left(- \cfrac{1}{2T(x/N)} (p_x^2 + (r_x -\ell)^2) \right)$$
where $T(\cdot)$ is the linear interpolation on $[0,1]$ between $T_\ell$ and $T_r$. ${\tilde \mu}$ is the "local equilibrium" approximation of $<\cdot>_{ss}$. Let $f_{ss}$ be the density of the stationary state $<\cdot>_{ss}$ with respect to ${\tilde \mu}$, i.e. the solution of ${\mathcal L}^{*,T(\cdot)} f_{ss} =0$. Here ${\mathcal L}^{*,T(\cdot)}$ is the adjoint operator of ${\mathcal L}$ in ${\mathbb L}^2 ({\tilde \mu})$. It turns out that
\begin{eqnarray*}
{\mathcal L}^{*,T(\cdot)}&=& -{\mathcal A} +\gamma {\mathcal S} + B_{1,T_\ell} +B_{N,T_r} \\
&+&\cfrac{\delta T}{T^2}   \left( \cfrac{1}{N} \sum_{x=1}^{N-2} {\tilde j}^{e}_{x,x+1} -\ell \, \cfrac{1}{N} \sum_{x=1}^{N-1} {\tilde j}^r_{x,x+1}\right)\\
&+&\cfrac{\delta T}{T^2} \left( \cfrac{1}{N} \sum_{x=1}^{N-1} p_x p_{x+1} X_{x,x+1}\right) +\cfrac{\delta T}{4} (\partial^2_{p_1} -\partial^{2}_{p_N})\\
&+& {\mathcal O} ((\delta T)^2) +{\mathcal O}(N^{-1})
\end{eqnarray*}
where ${\hat j}^{e}$ and ${\hat j}^r$ are the energy and deformation currents for the reversed dynamics at equilibrium. They are obtained from $j^e$ and $j^r$ by reversing momenta $p \to -p$. Expanding $f_{ss}$ at first order $f_{ss}= 1 +\delta T\, v + o(\delta T)$, we get that for large $N$ and small gradient temperature $\delta T$, $v$ has to satisfy the following Poisson equation:
\begin{equation*}
(-{\mathcal A} + \gamma {\mathcal S}) v = T^{-2} \left(\cfrac{1}{N} \sum_{x=1}^{N-2} {\tilde j}^{e}_{x} -\ell \, \cfrac{1}{N} \sum_{x=1}^{N-1} {\tilde j}^r_{x}\right)
\end{equation*} 
Let ${\hat v}$ the function obtained from $v$ by reversing momenta. By the fluctuation-dissipation equation (\ref{eq:fde}) we get
\begin{equation*}
{\hat v}=\cfrac{1}{NT^2}\sum_{x=1}^{N-1} (g_x- \ell h_x ) + {\mathcal O} (N^{-1})
\end{equation*} 
Therefore the functions $g_x$ and $h_x$ are directly related to first order corrections to local equilibrium. 
 
\section{{\label{sec:5}} Non-equilibrium Fluctuations and steady State Correlations}

Assume that initially the system starts from a local equilibrium $<\cdot>$ with macroscopic deformation profile $u_0 (q)$ and temperature profile $T_0 (q)$, $q \in [0,1]$. The time-dependant deformation fluctuation field $R_t^N$ and energy fluctuation field $Y_t^N$ are defined by
$$R_t^N (H)=\cfrac{1}{\sqrt N}\sum_{x=1}^{N} H\left (x/N \right)\left(r_x (t) - u\left(t,x/N\right) \right)$$    
$$Y_t^N (G)=\cfrac{1}{\sqrt N}\sum_{x=1}^{N} G\left(x/N\right)\left(e_x (t) -\varepsilon(t,x/N) \right)$$
where $H,G$ are smooth test functions, $(T(t,\cdot), u(t,\cdot))$ are solutions of the hydrodynamic equations (\ref{eq:hl}) with $\varepsilon=T+u^2 /2$. 

The fluctuation-dissipation equations (section \ref{SEC:3}) give (\cite{FNO}, \cite{S1}):
\begin{equation*}
\begin{cases}
R_t^N (H)=R_0^N (H)+ \cfrac{1}{\gamma}\int_{0}^t R_s^N(\Delta  H)ds+{\cM}_t^{1,N}\\
Y_t^N (G)=Y_0^N (G)+\gamma \int_{0}^{t} Y_s^N (\Delta G)ds\\
\phantom{Y_t^N (G)} + \int_{0}^{t}ds\left\{\cfrac{1}{\sqrt{N}}\sum_{x\in \TT_N}(\Delta G)(x/N)f_x(\omega_s)\right\}\\
\phantom{Y_t^N (G)} +{\cM}_t^{2,N} 
\end{cases}
\end{equation*}
where ${\cM}^{1,N}$ and ${\cM}^{2,N}$ are martingales and $f_x$ is the function defined by
$$f_x (\omega)= \cfrac{\left(\gamma^{-1}-\gamma\right)}{2} r_x^2-\left(\cfrac{1}{2\gamma}p_{x+1}p_x -\cfrac{\gamma}{4}\nabla^{*} p_x^2\right)$$

Covariance of the limit martingales are computed using standard stochastic calculus and thermal equilibrium property (\cite{S1}, \cite{FNO}):
\begin{equation*}
 \left<\left({\cM}^{1,N}_t\right)^2\right> \rightarrow \cfrac{2}{\gamma}\int_0^t ds \int_{[0,1]}dq T(q,s)(\nabla H)^{2}(q)
\end{equation*}

\begin{equation*}
\begin{split}
 \left<\left({\cM}^{2,N}_t\right)^2\right> \rightarrow \cfrac{2}{\gamma}\int_{[0,1]}dq \int_0^t ds u^{2} (q,s) T(q,s)(\nabla G)^{2}(q)\\
+(\gamma+\gamma^{-1})\int_0^t ds \int_{[0,1]}dq T^{2}(q,s)(\nabla G)^{2}(q)
\end{split}
\end{equation*}

\begin{eqnarray*}
 \left<{\cM}^{1,N}_t {\cM}^{2,N}_t\right> \rightarrow \\
\cfrac{2}{\gamma}\int_0^t ds\int_{[0,1]}dq u(s,q) T(s,q)(\nabla G)(q)(\nabla H)(q)
\end{eqnarray*}

Hence $R_t^N$ converges as $N$ goes to infinity to the solution of the linear stochastic differential equation:
\begin{equation}
\label{eq:R}
\partial_t R =\cfrac{1}{\gamma}\Delta R -\nabla\left[\sqrt{\cfrac{2}{\gamma}T(t,q)}W_{1}(t,q)\right]
\end{equation}
where $W_1(t,q)$ is a standard space time white noise.\\

The description of the limit for the energy fluctuation field is more demanding. We have first to close the equation. In order to do it, we use a "dynamical Boltzmann-Gibbs lemma" (\cite{KL}, \cite{S1}). Observables are divided into two classes: non-hydrodynamical and hydrodynamical. The first one are non conserved quantities and fluctuate in a much faster scale than the others (in the time scale where these last change). Hence, they should average out and only their projection on the hydrodynamical variables should persist in the limit. One expects there exist constants $C, D$ such that  
\begin{eqnarray*}
\cfrac{1}{\sqrt N} \int_0^t ds \sum_{x=1}^{N} (\Delta G)(x/N) \left\{ f_x (\omega_s)\right.\\
\left. -C(r_x -u(s,x/N)) -D\left(\cE_x-\varepsilon (s,x/N)\right)\right\}
\end{eqnarray*}
vanishes as $N$ goes to infinity. Constants $C$ and $D$ depend on the macroscopic point $q=x/N$ and on the time $t$. In order to compute these constants, we assume thermal local equilibrium. Around the macroscopic point $q$, the system is considered in equilibrium with a fixed value of the deformation $u(t,q)$ and of the temperature $T(t,q)$. The constant $C,D$ are then computed by projecting the function $f_x$ on the deformation and energy fields (\cite{KL}, \cite{S1}). If $\mu^{T,\ell}$ is the Gibbs equilibrium measure with temperature $T$ and mean deformation $\ell$ (the mean energy is then ${\varepsilon}={\ell}^{2}/2 +T$), we have $\Phi(\ell,{\varepsilon})=\mu^{T,\ell}(f_{x})=\varepsilon +{\ell}^{2}/2$ and then
\begin{equation*}
C=\partial_\ell \Phi (u(s,q), {\varepsilon} (s,q)), \quad D= \partial_{\varepsilon} \Phi (u(s,q),{\varepsilon} (s,q))
\end{equation*}

Therefore the time-dependant energy fluctuation field $Y_t^N$ converges as $N$ goes to infinity to the solution of the linear stochastic differential equation:
\begin{widetext}
\begin{equation}
\label{eq:Y}
\partial_t Y =\cfrac{1}{2}\left(\gamma+\cfrac{1}{\gamma}\right)\Delta Y +\cfrac{1}{2}\left(\cfrac{1}{\gamma}-\gamma\right)\Delta (u(t,q) R)
-\nabla\left[\sqrt{\gamma +\gamma^{-1}}T(t,q)W_{2}(t,q)+u(t,q)\sqrt{\cfrac{2T(t,q)}{\gamma}}W_1(t,q)\right]
\end{equation}
\end{widetext}
where $W_2 (t,q)$ is a standard space-time white noise independent of $W_1 (t,q)$. 

Remark that the deterministic terms in (\ref{eq:R}) and (\ref{eq:Y}) result from linearizing the nonlinear equation as (\ref{eq:hl}).

We now compute the fluctuations fields for the NESS $<\cdot>_{ss}$ which is obtained as the stationary solution of the Langevin equations (\ref{eq:R}-\ref{eq:Y}).  The field $L_t$ defined by $L_t= -\ell R_t +Y_t$ is solution of the Langevin equation
\begin{equation*}
\partial_t L = b\Delta L -\nabla \left[\sqrt{2b}{T} (t,q) W_2 (q,t) \right]
\end{equation*} 
with $b=\cfrac{1}{2} (\gamma +\gamma^{-1})$. The fields $R_t$ and $L_t$ are solutions of independent decoupled linear Langevin equations and converge as $t$ goes to infinity to independent Gaussian fields. It follows that $R_t$ and $L_t$ converge to stationary fluctuation fields ${R}_{ss}$ and ${Y}_{ss}$ such that
\begin{eqnarray*}
\text{Cov}({R}_{ss}(G),{R}_{ss}(H))=\int_{0}^{1}dq G(q)H(q){\bar T}(q)\\
\text{Cov}({Y}_{ss}(G),{Y}_{ss} (H))=\int_{0}^{1}dq G(q) H(q)\left\{{\bar T}^2 (q)+\ell^2{\bar T}(q)\right\}\\
+2(T_\ell -T_r)^{2} \int_0^1 G(q) (\Delta^{-1} H) (q) dq\\
\text{Cov}({Y}_{ss} (G),{R}_{ss}(H))=\ell\int_{0}^{1} H(q) G(q){\bar T}(q)dq
\end{eqnarray*}

Observe that the covariance of the fluctuations of energy is composed of two terms. The first one corresponds to Gaussian fluctuations for the energy under local equilibrium state while the second term represents the contribution to the covariance due to the long range correlations in the NESS. As in the case of SSEP and KMP process, the correction is given by the Green function of the Dirichlet Laplacian (\cite{BL3}, \cite{S2}).

\section{{\label{sec:6}} Large fluctuations}

\subsection{Macroscopic dynamical behavior}
Assume that initially the system is prepared in a state with a deformation profile $u_0$, energy profile ${\varepsilon}_0$ and hence temperature profile $T_0= {\varepsilon}_0 -{u}_0^2 /2$. In a diffusive scale the deformation (resp. energy, resp. temperature) profiles $u$ (resp. ${\varepsilon}$, resp. $T$) where ${\varepsilon}=T+{u^2}/2$ evolve according to the hydrodynamic equations (\ref{eq:hl}). 

Our aim is to obtain the large deviation principle corresponding to the law of large numbers $(\ref{eq:hl})$. It consists to estimate the probability that the empirical quantities (deformation, energy, temperature) do not follow the corresponding solutions of $(\ref{eq:hl})$ but remain close to some prescribed paths. This probability will be exponentially small in $N$ and we look for the exponential rate. We follow the classic procedure in large deviation theory (\cite{BL2},\cite{KL}): we perturb the dynamics in such a way that the prescribed paths become typical and we compute the cost of such perturbation.\\

Fix a path ${\mathcal Y} (t, \cdot)=(u(t,\cdot),{\varepsilon} (t,\cdot))$.  The empirical deformation profile ${\mathcal R}^N_t$ and empirical energy profile ${\mathcal E}_t^N$ are defined by
\begin{eqnarray}
\label{eq:RE0}
{\mathcal R}_t^N (q)= N^{-1} \sum_{x=1}^N r_x (t) {\bf 1}_{\left[x/N, (x+1)/N \right)} (q),\\
{\mathcal E}_t^N (q)= N^{-1} \sum_{x=1}^N e_x (t) {\bf 1}_{\left[x/N, (x+1)/N \right)} (q).\nonumber 
\end{eqnarray}
In appendix, we explain how to define a Markovian dynamics associated to a couple of functions $H(t,q), G(t,q)$, $q \in [0,1]$, such that the perturbed system has hydrodynamic limits given by $u$ and $\varepsilon$. This is possible if the function $F=(H,G)$ solves the Poisson equation
 \begin{equation}
\label{eq:poisson} 
\begin{cases}
\partial_t {\mathcal Y}= \Delta {\mathcal Y} -\nabla(\sigma \nabla F))\\ 
F(t,0)=F(t,1)=(0,0)
\end{cases}
\end{equation}
where the mobility $\sigma:=\sigma (u,\varepsilon)$ is given by 
\begin{equation}
\label{eq:mobility}
\sigma(u,\varepsilon)=
2\left(
\begin{array}{cc}
T & uT\\
uT & u^2 T +T^2
\end{array}
\right), \quad T={\varepsilon}-u^2 /2
\end{equation}

The perturbed process defined a probability measure $\tilde{\mathbb P}$ on the deformation/energy paths space by mean of the empirical deformation and energy profiles (see (\ref{eq:RE0})).

Our goal is to estimate the probability
\begin{eqnarray*}
\phantom{a}&\phantom{=}&\PP \left[({\mathcal R}_s^N, {\mathcal E}_{s}^N)\sim(u(s, \cdot),\varepsilon (s, \cdot)),\; s\in [0,t]\right]\\
&=&{\tilde \EE} \left[ \cfrac{d\PP}{d{\tilde{\PP}}} {\bf 1}_{\left\{({\mathcal R}_s^N, {\mathcal E}_{s}^N )\sim(r(s, \cdot),\varepsilon (s, \cdot)), \; s\in[0,t]\right\}}\right]
\end{eqnarray*}

To avoid irrelevant complications due to the fluctuations of the initial state which have no incidence on the derivation of the quasi-potential, we assume that the initial profiles $u_0$ and $T_0$ are the stationary profiles ${\bar r} (q)=\ell$ and ${\bar T} (q) =T_\ell +(T_r -T_\ell)q$. The function $F$ is such that 
$${\tilde \PP}  \left[({\mathcal R}_s^N, {\mathcal E}_{s}^N)\sim(u(s,\cdot),{\varepsilon} (s, \cdot)),\; s \in [0,t]\right]\approx 1$$
In the appendix we show that in the large $N$ limit, under ${\tilde {\mathbb P}}$, the Radon-Nikodym derivative is given by
\begin{equation*}
\cfrac{d\PP}{d{\tilde{\PP}}} \approx \exp\left\{ -N J_{[0,t]}(u,{\varepsilon})\right\}
\end{equation*} 
where
\begin{equation}
\label{eq:29}
J_{[0,t]} (u,{\varepsilon})=\cfrac{1}{2}\int_0^{t} ds<\nabla F (s, \cdot), \sigma \nabla F (s,\cdot)>_q
\end{equation}
where $\sigma$ is here for $\sigma (u(s,\cdot),\varepsilon(s,\cdot))$ and $<\cdot,\cdot>_q$ for the usual scalar product in ${\mathbb L}^2 ([0,1],dq)$. Hence we have obtained
\begin{eqnarray*}
\PP \left[({\mathcal R}_s^N, {\mathcal E}_{s}^N)\sim(u(s, \cdot),{\varepsilon}(s, \cdot)),\; s\in [0,t]\right] \\
\approx \exp\left\{ -N J_{[0,t]}(u,{\varepsilon})\right\}
\end{eqnarray*}

\subsection{The quasi-potential}

To understand what is the quasi-potential, consider the following situation. Assume the system is macroscopically in the stationary profile $(u(-\infty, \cdot), {\varepsilon} (-\infty,\cdot))= (\ell, {\bar T}(\cdot)+\ell^2/2)$ at $t=-\infty$ but at $t=0$ we find it in the state $(u(q),\varepsilon(q))$. We want to determine the most probable trajectory followed in the spontaneous creation of this fluctuation. According to the precedent subsection this trajectory is the one that minimizes $J_{[-\infty,0]}$ among all trajectories $({\hat u}, {\hat \varepsilon})$ connecting the stationary profiles to $(u,\varepsilon)$. The quasi-potential is then defined by
\begin{equation*}
W (u,\varepsilon) = \inf_{({\hat u}, {\hat \varepsilon})} J_{[0,t]} ({\hat u}, {\hat \varepsilon})
\end{equation*} 
MFT postulates the quasi-potential $W$ is the appropriate generalization of the free energy for non-equilibrium systems and this has been proven rigorously for SSEP (\cite{BL2}). $W$ is solution of an infinite-dimensional Hamilton-Jacobi equation which is in general very difficult to solve. It has been solved for specific models (SSEP and KMP) having a single conservation law (\cite{BL2}, \cite{BL3}). For the system we consider, two quantities are conserved and  we are not able to solve this Hamilton-Jacobi equation. Nevertheless we can compute the quasi-potential for the temperature profile (\ref{eq:30}) in the case $\gamma=1$. The latter is obtained by projecting the quasi-potential $W$ on the deformation/energy profiles with a prescribed temperature profile. 

Consider the system in its steady state $<\cdot>_{ss}$. Our aim is here to estimate the probability that the empirical kinetic energy defined by
\begin{equation}
\label{eq:30}
\Theta^N (q) = N^{-1} \sum_{x=1}^N p_x^2 {\bf 1}_{\left[x/N, (x+1)/N \right)} (q)
\end{equation}
is close to some prescribed temperature profile $\pi (q)$ different form the linear profile ${\bar T} (q)= T_\ell +(T_r -T_\ell)q$. This probability will be exponentially small in $N$ 
\begin{equation*}
\left< \left[ \Theta^N (q) \sim \pi (q) \right] \right>_{ss} \approx \exp( -N V(\pi))
\end{equation*}
By MFT, the rate function $V(\pi)$ coincides with the following \textit{projected quasi-potential}
\begin{equation*}
V(\pi)=\inf_{t >0} \inf_{(u,\varepsilon)\in \cA_{t,\pi}} J_{[0,t]} (u,\varepsilon)
\end{equation*}
where the paths set ${\cA}_{t,\pi}$ is defined by
\begin{eqnarray*}
{\cA}_{t,\pi} = \left\{ (u,\varepsilon); \quad {\varepsilon}(t, \cdot)-\cfrac{u^2(t, \cdot)}{2}=\pi(\cdot); \right.\\
\left. {\phantom{\cfrac{u^2(t, \cdot)}{2}}}u (0,\cdot)=\ell,\, T(0,\cdot)= {\bar T}(\cdot)\right\}
\end{eqnarray*}
Paths ${\mathcal Y}=(u,\varepsilon) \in {\mathcal A}_{t,\pi}$ must also satisfy the boundary conditions 
\begin{equation}
\label{eq:bc}
u(t,0)=u(t,1)=\ell, \quad \varepsilon (t,0)=T_\ell+\ell^2/2, \, \varepsilon (t,1)=T_r +\ell^2/2
\end{equation}
In fact, it can be shown that $J_{[0,t]} (u,\varepsilon) =+\infty$ if the path ${\mathcal Y}$ does not satisfy these boundary conditions.\\

Our main result is the computation of the projected quasi-potential:
\begin{equation}
\label{eq:32}
V (\pi)= \inf_{\tau \in \cT} [{\cF} (\pi, \tau) ]
\end{equation}
where $\cT=\{ \tau \in C^{1}([0,1]); \; \tau' (q)>0,\; \tau(0)=T_\ell,\; \tau (1)=T_r\}$ and 
\begin{equation*}
{\cal F} (\pi,\tau)= \int_0^1 dq \left[\cfrac{\pi (q)}{\tau(q)} -1 - \log\cfrac{\pi(q)}{\tau (q)} -\log \cfrac{\tau' (q)}{(T_r - T_{\ell})}\right]
\end{equation*}

Before proving (\ref{eq:32}) let us make some remarks. First, $V(\pi)$ is equal to the rate function for the KMP process (\cite{BL3}). Nevertheless, it is not easy to understand the deep reason. The symmetric part ${\mathcal S}$ of the generator ${\mathcal L}$ is more or less a time-continuous version of the KMP process for the kinetic energy but the Hamiltonian part has a non-trivial effect on the latter since it mixes momenta with positions. Hence, the derivation of the quasi-potential for the kinetic energy can not be derived from the the computations for the KMP process. Secondly, we are able to compute $V$ only for $\gamma=1$. When $\gamma$ is equal to $1$ hydrodynamic equations for the deformation and for the energy are decoupled but since temperature is a non-linear function of deformation and energy, it is not clear why it helps-- but it does. 
Formula (\ref{eq:32}) shows that the large deviation functional $V$ is nonlocal and consequently not additive: the probability of temperature profile in disjoint macroscopic regions is not given by the product of the separate probabilities. Nonlocality is a generic feature of NESS and is related to the ${\mathcal O} (N^{-1})$ corrections to local thermal equilibrium. 

\vspace{0,5cm}
Let us call $S(\pi)$ the right hand side of equality (\ref{eq:32}). 
\vspace{0,5cm}

For every time independent deformation/energy profiles $(r(q),e(q))$  and $\tau (q) \in \cT$ we define the functional 
\begin{equation}
\label{eq:U}
U(r,e,\tau)=\int_0^1 dq \left\{ \cfrac{T}{\tau} -1 - \log \cfrac{T}{\tau} -\log \cfrac{\tau'}{T_\ell - T_r} +\cfrac{(r-\ell)^2}{2\tau}\right\}
\end{equation}  
where $T(q)=e(q)-r(q)^2/2$ the temperature profile corresponding to $(r(q),e(q))$. Define the function $\tau:=\tau(r,e)$ of ${\cT}$ as the unique increasing solution of:
\begin{equation}
\label{eq:tau}
\begin{cases}
\tau^2 \cfrac{\Delta \tau}{(\nabla \tau)^2} = \tau -T -\cfrac{1}{2} (r-\ell)^2\\
\tau(0)=T_\ell,\; \tau (1)= T_r
\end{cases}
\end{equation}
Fix deformation/energy paths satisfying boundary conditions (\ref{eq:bc}) and define ${\mathcal Z}$ by
\begin{equation}
\label{eq:YZ}
{\mathcal Y}=\left(\begin{array}{c}u\\\varepsilon \end{array}\right),\qquad {\mathcal Z}=\left[\partial_t {\mathcal Y}-\Delta {\mathcal Y}+\nabla(\sigma \nabla(\delta U))\right],\\
\end{equation}
In the appendix we show the following formula
\begin{widetext}
\begin{eqnarray}
\label{eq:43}
J_{[0,t]}(u,\varepsilon)=U(u(t,\cdot),{\varepsilon} (t,\cdot),\tau({\varepsilon} (t,\cdot),u(t,\cdot)))-U(u(0, \cdot),{\varepsilon}(0,\cdot),\tau(u(0,\cdot),{\varepsilon} (0,\cdot)))\\
\nonumber \\
+\cfrac{1}{2}\int_0^t ds \left<\nabla^{-1}{\mathcal Z}, \sigma^{-1}\nabla^{-1}{\mathcal Z} \right>_q+\cfrac{1}{4}\int_0^t ds  \int_0^1 dq (u(s,q)-\ell)^4 \cfrac{(\nabla \tau)^2 (s,q)}{\tau^4 (s,q)}\nonumber
\end{eqnarray}
\end{widetext}
where 
\begin{equation*}
\delta U=\left(\begin{array}{c}\cfrac{\delta U}{\delta r}\\ \phantom{a}\\ \cfrac{\delta U}{\delta {e}} \end{array}\right) (u,\varepsilon,\tau(u,\varepsilon))
\end{equation*}

If $(u,\varepsilon)$ belongs to ${\mathcal A}_{t,\pi}$, 
\begin{eqnarray*}
U(\, u(0, \cdot),{\varepsilon}(0,\cdot),\tau(u(0,\cdot),{\varepsilon} (0,\cdot))\,)\\
=U(\,\ell, \bar{T}+\ell^2/2,\tau(\ell, {\bar T} +\ell^2 /2)\,)=0
\end{eqnarray*}
and 
$$U(u(t, \cdot),{\varepsilon} (t,\cdot), \tau(u(t, \cdot), \varepsilon(t, \cdot))) \ge {\mathcal F} (\pi, \tau (u(t,\cdot),{\varepsilon} (t,\cdot)).$$
The two last terms on the right hand side of (\ref{eq:43}) are positive so that for every paths in ${\mathcal A}_{t,\pi}$, we have
\begin{eqnarray*}
J_{[0,t]}(u,\varepsilon) \geq S(\pi)
\end{eqnarray*}
and we obtain hence 
\begin{equation}
\label{eq:firstinequality}
V(\pi)\geq S(\pi)
\end{equation}

To obtain the other sense of the inequality, we have to construct an optimal path $(u^*,{\varepsilon}^*) \in \cA_{t,\pi}$ such that the two last terms in the right hand side of $(\ref{eq:43})$ are equal to $0$, i.e.
\begin{equation}
\label{eq:44}
\begin{cases}
\partial_t {\mathcal Y}=\Delta {\mathcal Y}-\nabla(\sigma \nabla(\delta U))\\ 
u(t,q)=\ell
\end{cases}
\end{equation}
We note $T^{*}={\varepsilon}^* -{u^*}^2/2$ the corresponding temperature. Then reporting in (\ref{eq:43}), we obtain
\begin{equation}
\label{eq:10000}
J_{[0,t]}(u^*,{\varepsilon}^*)=U(u^{*}(t),{\varepsilon}^{*}(t),\tau(u^*(t) , {\varepsilon}^*(t)))  
\end{equation}
By the definition (\ref{eq:U}) of $U$ and by using the fact that $u^{*} (t,q)=\ell$, we obtain
\begin{eqnarray}
\label{eq:10001}
J_{[0,t]}(u^*,{\varepsilon}^*)={\cF}(T^{*}(t, \cdot),\tau(u^*(t, \cdot), {\varepsilon}^{*}(t, \cdot)) \nonumber\\
={\cF}(\pi,\tau(\ell,\pi +\ell^2 /2))  
\end{eqnarray}
The variational problem defining $S$ is solved for $\tau= \tau(\ell,\pi+\ell^2 /2)$ (\cite{BL3}) so that 
$$S(\pi)={\cF} (\pi, \tau(\ell,\pi +\ell^2 /2))$$     
and therefore we have 
\begin{equation*}
V(\pi)=\inf_{t >0} \inf_{\cA_{t,\pi}} J_{[0,t]}(u,\varepsilon) \leq S(\pi) 
\end{equation*}
This inequality with (\ref{eq:firstinequality}) shows that $V(\pi)=S(\pi)$. 
It remains to prove that such ``good'' path exists. The proof is similar to \cite{BL3} and we shall merely outline it. Equation (\ref{eq:44}) is equivalent to the following one
\begin{equation}
\label{eq:45}
\begin{cases}
\begin{array}{l}
\partial_t T^{*}=-\Delta(T^*)+2\nabla\left[\cfrac{(T^*)^2}{(\tau^*)^2}\nabla(\tau^*)\right]\\ 
u^*(t,q)=\ell
\end{array}
\end{cases}
\end{equation}
where $\tau^*(t,\cdot)= \tau(\ell, T^*(t,\cdot) +\ell^2/2)$. Let us denote by $\theta^*(s, \cdot)=T^{*}(t-s, \cdot)$ the time reversed path of $T^*$. $\theta^{*}$ can be constructed in the following procedure. We define $\theta^{*}(s,q),\; s\in[0,t],\; q\in [0,1]$ by
$$\theta^{*}(s, \cdot)= \rho (s, \cdot) -2\rho(s, \cdot)^2 \cfrac{\Delta \rho (s, \cdot)}{[(\nabla \rho)(s, \cdot)]^2}$$ 
where $\rho(s,q)$ is the solution of
\begin{equation*}
\label{eq47}
\begin{cases}
\partial_s \rho =\Delta \rho\\
\rho (s,0)=T_{\ell},\qquad \rho (s,1)=T_r\\
\rho(0,q)=\rho_0(q)=\tau(\ell, \pi+\ell^2/2)(q)
\end{cases}
\end{equation*}
It can be checked that $T^{*}(s,q)=\theta^{*}(t-s,q)$ solves $(\ref{eq:45})$. Moreover, we have $T^{*}(0,\cdot)=\theta^{*}(t,\cdot)$ and $T^{*}(t,\cdot)=\pi(\cdot)$. This path belongs to ${\cA}_{t,\pi}$ only as $t\to \infty$ since $\theta^{*}(t,\cdot)$ goes to $\bar{T}(\cdot)$ as $t\to +\infty$. We have hence in fact to define $T^*$ by the preceding procedure in some time interval $[t_1,t]$ and to interpolate $\bar{T}(\cdot)$ to $\pi^{*}(t_1, \cdot)$ in the time interval $[0,t_1]$ (see \cite{BL2}, \cite{BL3} for details). 
This optimal path is also obtained as the time reversed solution of the hydrodynamic equation corresponding to the process with generator ${\cL}^{*}$. It is easy to show that this last hydrodynamic equation is in fact the same as the hydrodynamic equation corresponding to ${\cL}$. This is the "generalized" Onsager-Machlup theory developed in \cite{BL1} for NESS: "the spontaneous emergence of a macroscopic fluctuation takes place most likely following a trajectory which can be characterized in terms of the time reversed process."  Observe also the following a priori non trivial fact: the optimal path is obtained with a constant deformation profile.     

\section{Conclusions}

In the present work we obtained hydrodynamic limits, Gaussian fluctuations and (partially) large fluctuations for a model of harmonic oscillators perturbed by a conservative noise. Up to now MFT has been restricted to gradient systems with a single conservation law. This work is hence the first one where MFT is applied for a non-gradient model with two conserved quantities. The quasi-potential for the temperature has been computed in the case $\gamma=1$ and it turns out that it coincides with the one of the KMP process. Our results show this system exhibits generic features of non-equilibrium models : long range correlations and non-locality of the quasi-potential.

Nevertheless our study is not completely satisfactory. It would be interesting to extend the previous results to the case $\gamma \neq 1$ and to compute the quasi-potential for the two conserved quantities and not only for the temperature. The difficulty is that there does not exist general strategy to solve the corresponding infinite-dimensional Hamilton-Jacobi equation.

\appendix
\section{The perturbed system}
Let us denote by $\omega (s)  = (p_x (s),r_x (s))_x$ the configuration of the process at time $s$ and ${\mathbb P}$ the probability measure on the deformation/energy paths up to time $t$ that the process $(\omega(s))_{0 \le s \le t}$ defines. It is well known (\cite{RY}) that if $M$ is a ${\mathbb P}$-martingale with quadratic variation $\left[ M \right]_t$ then the probability measure $\tilde{\mathbb P}$ with Radon-Nykodim derivative given by
\begin{equation}
\label{eq:rn}
\cfrac{d{\mathbb P}}{d{\tilde{\mathbb P}}} = \exp \left( M_t -\cfrac{1}{2} \left[ M \right]_t \right)
\end{equation} 
defines a Markov process $({\tilde \omega} (s))_{0 \le s \le t}$. In particular for a time dependant function $f(t,\omega)$ on the configuration space, the process
$$M_t =f(t,\omega(t))-f(0,\omega(0))-\int _{0}^{t} ds (\partial_s +N^2 {\mathcal L})f\, (s, \omega (s))ds$$
is a ${\mathbb P}$-martingale with quadratic variation $\left[ M \right]_t$ given by
\begin{equation}
\label{eq:qv}
\left[ M \right]_t= N^2 \int_{0}^t [\Gamma (f,f)](\omega (s)) ds
\end{equation}
where the "carr\'e du champ" operator $\Gamma (\cdot,\cdot)$ is defined by
\begin{eqnarray*}
\Gamma (u,v)&=&{\mathcal L} (uv)-u{\mathcal L}v- v {\mathcal L} u\\
&=& \sum_{x=1}^{N-1} X_{x,x+1} (u) X_{x,x+1} (v)\\
&+& T_{\ell} (\partial_{p_1} u) (\partial_{p_1} v)+ T_{r} (\partial_{p_N} u) (\partial_{p_N} v)
\end{eqnarray*}

The Markov process $\tilde \omega$ has then generator given by 
\begin{equation}
\label{eq:genpert}
{\tilde {\mathcal L}}= {\mathcal L} +\Gamma (f,\cdot)
\end{equation}

The first order partial differential operator $\Gamma (f,\cdot)$ can be seen as a perturbative drift.

\subsection{Hydrodynamic limit}

Fix a couple of smooth functions $H(t,q), G(t,q)$ vanishing at the boundaries and define
\begin{eqnarray*}
f&=&f_H +f_G \\
&=& \sum_{x=2}^{N-2} H(t,x/N) (r_x +\nabla h_x) + \sum_{x=2}^{N-2} G(t,x/N) (e_x+\nabla g_x)
\end{eqnarray*}
where $h_x, g_x$ are the functions appearing in the fluctuation-dissipation equation (\ref{eq:fde}).

By taylor expansion, we get
\begin{equation*}
X_{z,z+1}(f_H)= -\cfrac{1}{N} (\nabla H)(t,z/N) (p_{z+1} -p_z) +\mathcal O (N^{-2})
\end{equation*}  
and
\begin{equation*}
\begin{split}
X_{z,z+1} (f_G)= -\cfrac{1}{N} (\nabla G) (t,z/N) \theta_z +{\mathcal O} (N^{-2})
\end{split}
\end{equation*}
where $\theta_z$ is defined by
\begin{equation*}
\theta_z=  p_z p_{z+1}+\cfrac{1}{2}\left( p_{z+1}(r_z +r_{z-1})-p_z (r_{z+1}+r_z)\right) 
\end{equation*}

By (\ref{eq:fde}) and (\ref{eq:genpert}), the instantaneous deformation (resp. energy) current ${\tilde j}^r_{x-1,x}$ (resp. ${\tilde j}^e_{x-1,x}$) for the perturbed system is now
\begin{equation*}
\begin{cases}
{\tilde j}_{x}^r=-\nabla(r_x) +{\tilde{\mathcal L}}h_x -\Gamma (f,h_x)\\
(\text{resp. \,} {\tilde j}_{x}^e = \nabla \phi_x +{\tilde{\mathcal L}}g_x -\Gamma (f,g_x) -p_{x}p_{x-1} X_{x-1,x}(f) \,)
\end{cases}
\end{equation*}
In comparison with  fluctuation-dissipation equation (\ref{eq:fde}), currents are modified by terms of order ${\mathcal O} (N^{-1})$ (see below). 

By Taylor expansions, one has, up to ${\mathcal O} (N^{-2})$ corrections,
\begin{eqnarray*}
\Gamma (f,h_x)&\approx&- \cfrac{1}{N}(\nabla H)(t,x/N) \left\{ p_{x+1} (p_{x+1}-p_x)\right. \\
&\phantom{=}&\phantom{\cfrac{1}{N}(\nabla H)(t,x/N)}\left.- p_{x-1} (p_x -p_{x-1})\right\}\\
&-&\cfrac{1}{N}(\nabla G)(t,x/N) \left\{ p_{x+1} \theta_x -p_{x-1} \theta_{x-1}\right\} 
\end{eqnarray*}
and
\begin{eqnarray*}
\Gamma (f,g_x)\approx -\cfrac{1}{2N} (\nabla H)(t,x/N) \left\{(r_x +r_{x-1})\right.\\
\left. (p_{x+1}^2 +p_{x-1}^2 -p_x p_{x+1} -p_x p_{x-1})\right\}\\
-\cfrac{1}{2N} (\nabla G) (t,x/N) \left\{(r_x +r_{x-1}) (\theta_x p_{x+1} -\theta_{x-1} p_{x-1})\right\} 
\end{eqnarray*}
and
\begin{eqnarray*}
p_{x-1}p_x X_{x-1,x} f &\approx&-\cfrac{1}{N} (\nabla H) (t,x/N) p_{x-1}p_x (p_x -p_{x-1})\\
&-&\cfrac{1}{N} (\nabla G) (t,x/N) \theta_{x-1} p_{x-1}p_x 
\end{eqnarray*}

\phantom{a\\}
To obtain hydrodynamic equations for the perturbed process $\tilde \omega$ we use thermal equilibrium property. Observe that at equilibrium under the Gibbs measure $\mu^{T,\ell}_N$ with mean deformation $\ell$ and temperature $T$, we have 
\begin{widetext}
\begin{equation*}
\left( 
\begin{array}{c}
\mu^{T,\ell} (\Gamma (f,h_x))\\
\mu^{T,\ell} (\Gamma (f,g_x)+ p_{x-1}p_x X_{x-1,x}f)
\end{array}
\right)
=-N^{-1} \sigma(\ell, T+\ell^2/2) 
\left( 
\begin{array}{c}
(\nabla H)(t,x/N)\\
(\nabla G)(t,x/N)
\end{array}
\right)
\end{equation*}
\end{widetext}
with the mobility matrix $\sigma$ defined in (\ref{eq:mobility}). We repeat the arguments of section \ref{SEC:3} and we get that the hydrodynamic limit of $\tilde{\omega}$ is given by 
\begin{equation*}
\partial_t {\mathcal Y}= \Delta {\mathcal Y} -\nabla(\sigma \nabla F))
\end{equation*}
with boundary conditions like in (\ref{eq:hl}). 

\subsection{The dynamical large deviations function}

Fix a path ${\mathcal Y} =(u,\varepsilon)$ and consider the perturbed process defined above with $F=(H,G)$ chosen according to (\ref{eq:poisson}). We show here that in the large $N$ limit
\begin{equation*}
{\tilde \EE} \left[ \cfrac{d\PP}{d{\tilde{\PP}}} {\bf 1}_{\left\{({\mathcal R}_s^N, {\mathcal E}_{s}^N )\sim(u (s, \cdot),\varepsilon (s, \cdot)), \; s\in[0,t]\right\}}\right] \approx e^{-N J_{[0,t]} (u,\varepsilon)}
\end{equation*}
with $J_{[0,t]} (u,{\varepsilon})$ defined in (\ref{eq:29}). By (\ref{eq:rn}) and (\ref{eq:qv}), the logarithm ${\mathcal J}_{[0,t]} (\omega)$ of the Radon-Nykodim derivative $d{\PP}/d{\tilde{\PP}}$ is given by 
\begin{equation*}
\begin{split}
{\mathcal J}_{[0,t]} (\omega)= f(t,\omega(t))-f(0,\omega(0))\\
-\int _{0}^{t} ds (\partial_s +N^2 {\tilde {\mathcal L}})f\, (s, \omega(s))ds + \cfrac{N^2}{2} \int_0^t [\Gamma (f,f)](\omega(s)) ds
\end{split}
\end{equation*}
Observe that
\begin{equation*}
{\tilde{\mathcal L}} (r_x +\nabla h_x)= \Delta r_x +\nabla (\Gamma (f,h_x))
\end{equation*}
and
\begin{equation*}
{\tilde L} (e_x +\nabla g_x)= \Delta \phi_x +\nabla (\Gamma (f,g_x) +p_{x-1} p_x X_{x-1,x} f)
\end{equation*}

Moreover the term $\Gamma(f,f)=\Gamma(f,f_H)+\Gamma(f,f_G)$ gives a contribution equal to
\begin{equation*}
\begin{split}
\Gamma(f,f)= \sum_{x=2}^{N-2} H(t,x/N) \nabla \Gamma (h_x,f)\\
+\sum_{x=2}^{N-2} G(t,x/N) \nabla \left[ \Gamma (g_x,f) +p_{x-1} p_x X_{x-1,x} f\right]
\end{split}
\end{equation*}

Recall that $F=(H,G)$ has been chosen such that
$${\tilde \PP} \left[ \left\{({\mathcal R}_s^N, {\mathcal E}_{s}^N )\sim(u(s, \cdot),\varepsilon (s, \cdot)), \; s\in[0,t]\right\}\right]=1$$
By local equilibrium statement, integration by parts and the precedent computations for $\Gamma (f,h_x), \Gamma (f,g_x)$ and $p_{x-1}p_x X_{x-1,x} f$, one has  
\begin{equation*}
\begin{split}
{\mathcal J}_{[0,t]} (\omega) \approx -\cfrac{N}{2} \int_0^t ds \langle \nabla F (s,\cdot), \sigma \nabla F (s, \cdot) \rangle_q
\end{split}
\end{equation*}
so that
\begin{equation*}
{\tilde \EE} \left[ \cfrac{d\PP}{d{\tilde{\PP}}} {\bf 1}_{\left\{({\mathcal R}_s^N, {\mathcal E}_{s}^N )\sim(u (s, \cdot),\varepsilon (s, \cdot)), \; s\in[0,t]\right\}}\right] \approx e^{-N J_{[0,t]} (u,\varepsilon)}
\end{equation*}

\section{Proof of formula (\ref{eq:43})}     
 
The goal is to express $J_{[0,t]}(u,\varepsilon)$ as the sum of $ U(u(t,\cdot)),\varepsilon(t,\cdot), \tau(\varepsilon (t, \cdot),u(t,\cdot)))-  U(u(0,\cdot)),\varepsilon(0,\cdot), \tau(\varepsilon (0, \cdot),u(0,\cdot)))$ and positive terms.  We recall that
\begin{equation}
\label{eq:calculus}
J_{[0,t]}(u,\varepsilon)=\cfrac{1}{2} \int_0^t ds \langle \nabla F (s, \cdot), \sigma \nabla F (s, \cdot) \rangle_q
\end{equation}
where $\sigma=\sigma(u(s,\cdot),\varepsilon(s,\cdot))$ is the mobility matrix defined in (\ref{eq:mobility}). By the definitions of ${\mathcal Y}$ and ${\mathcal Z}$ (see (\ref{eq:YZ})), we have
$$\nabla[\sigma\nabla F]= \Delta{\mathcal Y} -\partial_t {\mathcal Y}= \nabla (\sigma \nabla(\delta U))-{\mathcal Z}$$
Inserting this last expression in (\ref{eq:calculus}) we get
\begin{equation*}
\begin{split}
J_{[0,t]}(u,\varepsilon)=\cfrac{1}{2} \int_0^t ds <\nabla^{-1} \left[ \nabla (\sigma \nabla(\delta U))-{\mathcal Z}\right],\\
\sigma^{-1} \nabla^{-1}\left[ \nabla (\sigma \nabla(\delta U))-{\mathcal Z} \right]>_q
\end{split}
\end{equation*} 
We develop the expression and we get
\begin{eqnarray*}
J_{[0,t]}(u,\varepsilon)&=&\cfrac{1}{2}\int_0^t ds \left<\nabla^{-1}{\mathcal Z}, \sigma^{-1}\nabla^{-1}{\mathcal Z}\right>_q\\
&+&\cfrac{1}{2} \int_0^{t} ds <\sigma \nabla(\delta U), \nabla (\delta U)>_q\\
&-&\int_{0}^{t} ds <\nabla^{-1} {\mathcal Z}, \nabla (\delta U)>_q
\end{eqnarray*}
By integration by parts and because of the boundary conditions, the second term and third term on the right hand side are equal to
\begin{eqnarray*}
\int_0^{t} ds <\partial_s {\mathcal Y}, \delta U>_q -\cfrac{1}{2}\int_0^{t} ds <\sigma \nabla(\delta U), \nabla (\delta U)>_q\\
-\int_0^{t} ds <\Delta {\mathcal Y}, \delta U)>_q
\end{eqnarray*}
The first term is the integral of the time derivative of $s \to U(u(s,\cdot),\varepsilon(s,\cdot), \tau (u(s,\cdot),\varepsilon(s,\cdot))$ because
$$\cfrac{\delta U}{\delta \tau} (r,e,\tau(r,e))=0$$
Hence it is equal to $U(u(t,\cdot),\varepsilon(t,\cdot), \tau (u(t,\cdot),\varepsilon(t,\cdot))-U(u(0,\cdot),\varepsilon(0,\cdot), \tau (u(0,\cdot),\varepsilon(0,\cdot))$.

We develop the two other terms using the expression of $\delta U$. A simple computation shows
\begin{equation*}
\cfrac{\delta U}{\delta r}=r/T -\ell/\tau, \qquad \cfrac{\delta U}{\delta e}= \tau^{-1}-T^{-1} 
\end{equation*}

Hence we get
\begin{equation}
\label{eq:exp1}
\begin{split}
\cfrac{1}{2}\int_0^{t} ds <\sigma \nabla(\delta U), \nabla (\delta U)>\\
=\int_{0}^{t} ds \int_{[0,1]} dq \left\{\cfrac{(\nabla u)^2}{T}+\cfrac{(\nabla T)^2}{T^2} -2\cfrac{\nabla \tau \nabla T}{\tau^2} +T^2 \cfrac{(\nabla \tau)^2}{\tau ^4}\right.\\
\left. +2(\ell -u) \cfrac{\nabla u \nabla \tau}{\tau^2} +T (u-\ell)^2 \cfrac{(\nabla \tau)^2}{\tau^4}\right\} 
\end{split}
\end{equation}
For the term
\begin{equation}
\int_0^{t} ds <\Delta {\mathcal Y}, \delta U)>_q
\end{equation}
we perform an integration by parts and we obtain
\begin{eqnarray}
\label{eq:exp2}
-\int_{0}^{t} ds \int_{[0,1]} dq \left\{\cfrac{(\nabla u)^2}{T}+ \ell \cfrac{\nabla u \nabla \tau}{\tau^2} \right.\\
\left. +\cfrac{(\nabla T)^2}{T^2}-\cfrac{\nabla T \nabla \tau}{\tau^2} -u\cfrac{\nabla u \nabla \tau}{\tau^2}\right\} \nonumber
\end{eqnarray}

The sum of the expressions (\ref{eq:exp1}) and (\ref{eq:exp2}) is equal to
\begin{equation*}
\begin{split}
\int_0^{t} ds  \int_{[0,1]} dq \left\{-\cfrac{\nabla \tau \nabla T}{\tau^2} + T^2 \cfrac{(\nabla \tau)^2}{\tau^4}\right.\\
+ \left. (\ell -u) \cfrac{\nabla u \nabla \tau}{\tau^2} +T (\ell -u)^2 \cfrac{(\nabla \tau )^2}{\tau ^4} \right\} 
\end{split}
\end{equation*}

Remark now that by integration by parts,
\begin{eqnarray*}
\int_{[0,1]} dq \cfrac{\nabla \tau \nabla T}{\tau^2}=\int_{[0,1]} dq \nabla(T -\tau) \cfrac{\nabla \tau}{\tau^2} + \int_{[0,1]} dq \cfrac{(\nabla \tau)^2}{\tau^2}\\
=\int_{[0,1]} dq (T -\tau) \left(\cfrac{2(\nabla \tau)^2}{\tau^3} - \cfrac{\Delta \tau}{\tau^2}\right)+ \int_{[0,1]}dq \cfrac{(\nabla \tau)^2}{\tau^2} 
\end{eqnarray*}
and
\begin{equation*}
\begin{split}
\int_{[0,1]} dq (\ell -u) \cfrac{\nabla u \nabla \tau}{\tau ^2}=-\cfrac{1}{2}\int_{[0,1]} dq \nabla [(\ell -u)^2]\cfrac{\nabla \tau}{\tau^2}\\
=\cfrac{1}{2} \int_{[0,1]} dq (\ell -u)^2 \left\{\cfrac{\Delta \tau}{\tau^2} - 2 \cfrac{(\nabla \tau)^2}{\tau^3}\right\}
\end{split}
\end{equation*}
Collecting all these facts and using the equation defining $\tau$, we obtain (\ref{eq:43}).\\

\vspace{3cm}     
\bibliography{biblio-hcm-PRL2}
\end{document}